\documentstyle[prd,aps,floats,epsfig]{revtex}
\def\half{\textstyle{1\over2}}
\def\quarter{\textstyle{1\over4}}
\newcommand{\bc}{\begin{center}}
\newcommand{\ec}{\end{center}}
\newcommand{\be}{\begin{equation}}
\newcommand{\ee}{\end{equation}}
\newcommand{\bq}{\begin{eqnarray}}
\newcommand{\eq}{\end{eqnarray}}
\newcommand{\bsq}{\begin{mathletters}}
\newcommand{\esq}{\end{mathletters}}
\begin{document}
\preprint{DTP/99/49, gr-qc/9906107}
\draft
\renewcommand{\topfraction}{0.8}
 
\title{Vortices and black holes in dilatonic gravity}
\author{Caroline Santos\footnote{On leave from: Departamento de F\'\i sica 
da Faculdade de Ci\^encias da Universidade do Porto, Rua do Campo Alegre 687, 
4150-Porto, Portugal.}\footnote{E-mail: C.D.S.D.Silva@durham.ac.uk} and 
Ruth Gregory\footnote{E-mail: R.A.W.Gregory@durham.ac.uk}}
\address{Centre for Particle Theory, 
Durham University, South Road, Durham, DH1 3LE, U.K.
}
 
\date{\today}
 
\maketitle

\begin{abstract}
\noindent

We study analytically black holes pierced by a thin vortex in 
dilatonic gravity for an arbitrary coupling of the vortex to the
dilaton in an arbitrary frame.  We show that the horizon of the charged 
black hole supports the long-range fields of the Nielsen-Olesen 
vortex that can be considered as black hole hair for both 
massive and massless dilatons. We also prove that extremal
black holes exhibit a flux expulsion phenomenon for a sufficiently
thick vortex.  We consider the gravitational back-reaction
of the thin vortex on the spacetime geometry and dilaton, and
discuss under what circumstances the vortex can be used to 
smooth out the singularities in the dilatonic C-metrics.
The effect of the vortex on the massless dilaton is to generate 
an additional dilaton flux across the horizon. 

\end{abstract}
 
\vspace{5mm}
 
\pacs{PACS numbers: 04.40.-b, 11.27.+d \hfill DTP/99/49,
      gr-qc/9906107}


\section{\bf Introduction.}
 
The extrapolation of the black hole `no-hair' conjecture,
initially proposed by Ruffini and Wheeler \cite{RW}
and stating that a stationary black hole is uniquely determined
by its mass, electromagnetic charge and angular momentum,
to the stronger statement of `no dressing' of the horizon,
has been proven to be false \cite{5Chrusciel}, \cite{AGK}.
A common feature of such `counterexamples' is that they involve 
nontrivial topology of the matter fields.
In particular, in reference \cite{AGK}, it was shown that for
the Abelian-Higgs model
in Einstein gravity, (see \cite{nohair} for the relevant no hair theorems),
a Schwarzschild black hole
could indeed support long hair, namely, a $U(1)$ vortex, which could
either pierce, or end on the black hole horizon. This latter case is
particularly interesting as it provides a decay channel for the 
disintegration of otherwise stable topological vortices~\cite{split,GH,E2}.

It was also established in reference \cite{AGK} that the
gravitational effect of a vortex which is thin relative
to the Schwarzschild radius of the black hole
is to change its metric to
a smooth version of the Aryal, Ford and Vilenkin solution \cite{AFV}:
\be\label{afvmet}
ds^2 = \left(1-\frac{2E}{r} \right) dt^2
- \left(1-\frac{2E}{r}\right)^{-1} dr^2
- r^2 d\theta^2 - r^2 (1 - 4G\mu)^2 \sin^{2}\theta d\varphi^2
\ee
in which spacetime is asymptotically locally flat, but has
a conical deficit angle $8\pi G \mu$
for a string with energy density per unit length of $\mu$.
(For a string ending on a black hole, the metric was shown in 
\cite{GH} to be a smooth version of the C-metric \cite{KW}, or the 
Israel-Khan metric \cite{IK}, depending on whether a static or accelerating
black hole is required.)

This work was then extended to other black hole solutions,
namely to the Reissner-Nordstr\o m black hole in Einstein-Maxwell theory
\cite{CCES,BG,BRG} and to non-extreme electrically
charged black holes with a massless dilaton 
\cite{MR} in low energy string theory, and the main conclusion remains
the same, i.e., in the thin vortex limit the Abelian-Higgs vortex 
also provides hair for these black holes (although reference
\cite{MR} has an incorrect back reaction analysis).

In this paper we extend the work of reference \cite{AGK} to consider the
Abelian-Higgs model coupled to dilatonic gravity,
where the dilaton may be massless or massive.
Using the same method as in \cite{AGK} we show that a
Schwarzschild black hole can indeed support long hair, namely, a $U(1)$ vortex.
To lowest order the vortex (with an arbitrary  dilaton coupling ``$a$'')
introduces the same corrections on the geometry of the Schwarzschild black hole 
background  as in \cite{AGK}, and when the coupling of the dilaton to the vortex
is noncanonical in the string frame ($a \not= -1$),  
the vortex switches on non-constant values of the dilaton along the horizon.
We then extend these results to charged black holes, 
and using similar arguments we show that for a massless dilaton, 
magnetically charged black holes can support the Abelian-Higgs vortex 
for reasonable dilaton couplings to the vortex ($|a| \ll $O$(E^2)$).
For weak electrically charged black holes we prove analytically that 
they can support the Abelian-Higgs vortex, again for reasonable values of 
the dilaton couplings to the vortex.  To leading order,
the gravitational effect of the vortex on those black holes is to change their
background geometries in an analogous fashion to the AFV metric, namely
that a conical slice is removed from the geometry. However, for $a\neq-1$
the deficit angle is no longer constant, and acquires a dependence on the
background dilaton; in addition there are strong long range
gravitational effects to O($\epsilon^2$). The dilaton becomes modified
by a correction which has the same sign for both magnetic and electric black
holes, so that if its magnitude is decreased for the magnetic, it is
increased for the electric, and vice versa.

We also consider black holes with a massive dilaton which are qualitatively
different \cite{RJ,HH} from their massless cousins \cite{GHS}. As opposed
to the single horizon plus spacelike singularity causal structure of the 
massless dilatonic black holes, massive dilaton black holes can have two 
or three horizons and extremal solutions with
a double or triple degenerated horizon.

The layout of this paper is as follows: 
We first briefly review the self-gravitating
dilatonic $U(1)$ vortex in the next section. 
In section \ref{thinsec}\ we generalise to dilatonic strings the main results
of \cite{AGK} introducing the notations and the method. We examine the 
question of existence of the vortex in the Schwarzschild and charged
dilaton black hole backgrounds for a thin dilatonic 
cosmic string whether the dilaton is massless or massive.
In section \ref{gravsec}\ we study the gravitational effect of the
vortex, either in the background geometry or in the dilaton
and in particular on the horizon of those black holes.
Finally in section \ref{finsec}\ we summarise our results and conclude by
showing how to generalise the results to the case where the vortex is
coupled to the dilaton by an arbitrary parameter, $a$, in an arbitrary
frame, parametrised by $b$.

\section{\bf Dilatonic Strings.}\label{dilsec}
 
In this section we briefly review the self-gravitating 
dilatonic $U(1)$ vortex. This is based on the work in references \cite{GG,GS},
where we refer the reader for greater detail.
The  abelian-Higgs lagrangian is coupled  to the gravitational action 
in the string frame \cite{LESG} with an arbitrary coupling, ``$a$'', 
to the dilaton, $\phi$:
\be\label{saction}
{\hat S} = \int  d^4 x \sqrt{-{\hat g}} \left [ e^{-2\phi} \left ( -{\hat R}
- 4({\hat \nabla}\phi)^2 - {\hat V}(\phi) \right ) + e^{2a\phi} {\cal L}
\right ]
\ee
${\hat V}(\phi)$ represents a possible potential
for the dilaton and 
\be\label{abhiggs}
{\cal L}[\Phi ,B_c] = D_{a}\Phi ^{\dagger}D^{a}\Phi
- \frac{1}{4}{\tilde G}_{ab}{\tilde G}^{ab}
- \frac{\lambda}{4}(\Phi ^{\dagger} \Phi - \eta ^2)^2
\ee
is the abelian-Higgs lagrangian, where ${\tilde G}_{ab}$ is the field strength
of the gauge field $B_a$, whose mass in the broken phase $m_{v} =
\sqrt2 e \eta$ is related to the mass
of the Higgs field $\Phi$, $m_{H} = \sqrt\lambda \eta$,
by the Bogomolnyi parameter, $\beta = \frac{\lambda}{2 e^2}$ \cite{bog}.
(The self-gravitating Einstein vortex can be obtained 
by simply ignoring the dilaton.)
It is conventional to express the field content in a
slightly different manner in which the physical degrees of freedom are made
more manifest by defining real fields $X, \; \chi $ and $P_c$ by
\bsq
\begin{eqnarray}
\Phi (x^{\alpha}) &=& \eta X (x^{\alpha}) e^{i\chi(x^{\alpha}) }\\
B_{c} (x^{\alpha}) &=& \frac{1}{e} \left [ P_{c} (x^{\alpha}) 
- \nabla _{c} \chi (x^{\alpha}) \right ]
\end{eqnarray}
\esq
These fields represent the physical degrees of freedom of the broken symmetric
phase; $X$ is the scalar Higgs field, $P_c$ the massive vector boson
(with field strength $G=dP$), and
$\chi$, being a gauge degree of freedom, is not a local observable, but can
have a globally nontrivial phase factor which indicates the presence of a
vortex. The existence of vortex solutions in the abelian Higgs model was argued
by Nielsen and Olesen~\cite{NO}, and in the presence of a vortex $\oint d\chi =
2\pi N$, where $N$ is the winding number of the vortex.

The simplest vortex solution, and one which will form the basis of our 
analytic arguments, is that in flat space:
\be
X = X_0(R), \qquad P_\mu = NP_0(R)\partial_\mu\varphi,
\label{xpform}
\ee
where $R=r\sqrt{\lambda}\eta$, $\{r,\phi\}$ are
polar coordinates, and
$X_0$ and $P_0$ satisfy the coupled second order ODE's
\bsq \label{basic} \bq
-X_0'' - {X_0'\over R} + {X_0N^2P_0^2\over R^2} + {\half} X_0(X_0^2-1) &=& 0 \\
-P_0''+{P_0'\over R} + {X_0^2P_0\over\beta} &=& 0
\eq \esq
with prime denoting $d/dR$.
For $N=1$, this is the Nielsen-Olesen solution, and gives an isolated vortex
for all $\beta$. The vortex core consists of two components---a scalar core
where the Higgs field differs from vacuum, roughly of width $1/\sqrt{\lambda}
\eta$, and a gauge core of thickness O($\beta^{1/2}/\sqrt{\lambda} \eta$). For
higher $N$, the solutions were given in~\cite{AGHK}, the principal differences
to $N=1$ being that the $X$-field is flattened ($X\sim R^N$) near the core, and
the string is correspondingly fattened. An additional difference is that for
$\beta>1$, higher winding strings are unstable to separation into $N$ unit
winding vortices~\cite{bog}. 

For future reference we write the (normalised) energy-momentum tensor of the 
Nielsen-Olesen vortex:
\bsq\label{NOtens}
\bq
T^t_t = T^z_z = {\cal E}_0 (R)
&=& X_0 ^{\prime 2} +\frac{X_0^2 P_0^2}{R^2}
+\beta \frac{P_0 ^{\prime 2}}{R^2}+\frac{1}{4}(X_0 ^2 -1) ^2\\
T^R_R = - {\cal P}_{0R} (R) &=&
-X_0 ^{\prime 2} +\frac{X_0^2 P_0^2}{R^2}
-\beta \frac{P_0 ^{\prime 2}}{R^2}+\frac{1}{4}(X_0 ^2 -1) ^2 \\
T^\varphi_\varphi = - {\cal P}_{0\varphi} (R) &=&
X_0 ^{\prime 2} -\frac{X_0^2 P_0^2}{R^2}
-\beta \frac{P_0 ^{\prime 2}}{R^2}+\frac{1}{4}(X_0 ^2 -1) ^2
\eq
\esq
whose conservation law is 
\be\label{NOcons}
\left(R \, P_{0R}\right)^{\prime} = P_{0\varphi}.
\ee
 
Returning to the dilatonic vortex, it proves useful to write 
the action in terms of the ``Einstein'' metric, which is defined via    
\be\label{5tconforme}
g_{ab} = e^{-2\phi} {\hat g}_{ab}   
\ee
in which the gravitational part of the action appears in the
more familiar Einstein form:
\be\label{5eaction}
S = \int  d^4 x \sqrt{-g} \left [
- R + 2 (\nabla\phi)^2 - V(\phi)
+ 2\,\epsilon\,e^{2(a+2)\phi} 
{\hat {\cal L}} \{X,P,e^{2\phi}g \} \right ]
\ee
where $V(\phi) = e^{2\phi}\,{\hat V}$ and $\epsilon = \frac{\eta^2}{2}$ 
is the gravitational string coupling in these units.

For the cylindrically symmetric self-gravitating winding number one vortex,
a gauge can be chosen in which the matter fields take the
form (\ref{xpform}) to leading order, and the metric is
\be
ds_{cyl}^2 = e ^{\gamma}\left(dt ^2 - dR ^2 -dz ^2 \right)
 - \alpha^2 e ^{-\gamma } \, d\varphi^2 
\ee
where to order $\epsilon$ the geometry is
\bsq\bq
\alpha &=& \left [1- \epsilon \int _0 ^R
R ({\cal E}_{_0} - {\cal P}_{_0R})d R \right ] R +
\epsilon \int _0 ^R  R ^2 ({\cal E}_{_0} - {\cal P}_{_0 R})dR,
= [1-\epsilon A(R)]R + \epsilon B(R) \\
\gamma &=& \epsilon \int _0 ^R R {\cal P}_{_0 R} dR = \epsilon D(R) .
\eq\esq
which is actually identical to the form of the Einstein self-gravitating 
vortex. This metric asymptotes a conical spacetime with deficit angle
$2\pi \epsilon(A+D) = 2\pi\int_0^\infty R{\cal E}_0 dR = \epsilon\mu$, 
the characteristic signature of a cosmic string of energy per unit length
$\mu$ \cite{VHL}.  Note that in this, and what follows, 
we have chosen  ``vortex units'', in which the string 
width is of order unity (i.e. $\sqrt\lambda\eta = 1$).

For the dilaton, the equation of motion is
\be
(\alpha \phi')' = {\alpha e^\gamma\over4} {\partial
V\over\partial\phi} + \epsilon \alpha e^\gamma \left [ (a+1){\hat T}^t_t 
+ {\half} ({\hat T}^R_R + {\hat T}^\varphi_\varphi) \right ] 
\ee
where ${\hat T}_{ab}$ represents the energy-momentum tensor for the 
vortex fields, $X$ and $P$,
\be\label{stringem}
{\hat T}_{ab} = 2
e^{2(a+1)\phi} \left[ \nabla_aX\nabla_bX + X^2 P_aP_b \right] - \beta
e^{2a\phi} G_{ac}G_b^{\ c} - 
e^{2(a+2)\phi} {\hat {\cal L}} g_{ab} .
\ee
If $V(\phi)=0$, one obtains
\be\label{mslsdil}
\phi = -{\epsilon D(R)\over2} + \epsilon(a+1) \int_0^R {A+D\over R} 
\sim (a+1) {\epsilon \mu\over 2\pi} \ln R- {\epsilon D(\infty)\over2}
\;\;\;{\rm as} \;\; R\to\infty
\ee
This dilaton field has the effect that to O($\epsilon^2$), the geometry
acquires long range corrections, and on very large length scales is not
asymptotically locally flat.
The $a = -1$ massless dilatonic cosmic string has no long range
effects (other than the deficit angle) and merely shifts the value
of the dilaton between the core and infinity by a constant of
order $\epsilon$. For the special case $\beta = 1$, there is no effect
at all on the dilaton field \cite{GS}, and the 
dilatonic string is the same as the \rm Einstein one. 

For a massive dilaton assuming a potential $V(\phi) = 2M^2\phi^2$,
the general solution is instead
\bq
\phi &=& - \epsilon K_0(MR) \int_0^R I_0(MR') R' \left [ (a+1){\cal E}(R')
- {\half} ({\cal P}_R(R') + {\cal P}_\theta(R')) \right ] dR' \nonumber\\
&& \;\;\; - \epsilon I_0(MR) \int_R^\infty K_0(MR') R' \left [ (a+1){\cal
E}(R') - {\half} ({\cal P}_R(R') + {\cal P}_\theta(R')) \right ]  dR' \\
&\simeq &- (a+1) {\epsilon\mu\over 2\pi} K_0(MR) \;\;{\rm for}\;\;R>1,
\;\; M\ll1 \nonumber
\eq
where $K_0,I_0$ are the modified Bessel functions.
In the case that $a = -1$, the dilaton is very strongly damped to zero
outside the core therefore to a good approximation $\phi = 0$
outside the core, irrespective of $M$.

\section{\bf Strings through black holes.}\label{thinsec}

We now consider an isolated system of a dilatonic string threading
a black hole and argue the existence of a vortex 
solution in the absence of gravitational back reaction.
We begin by reviewing the argument of ref. \cite{AGK} for
the existence of a vortex solution in the Schwarzschild
black hole background:
\be\label{5MetricSch}
ds^2 = \left(1-\frac{2E}{r} \right) dt^2
- \left(1-\frac{2E}{r}\right)^{-1} dr^2
- r^2 \,(d\theta^2
+ \sin^{2}\theta \, d\varphi^2)
\ee
(where $E$ is the mass of the black hole 
measured in ``vortex units''), since this is also a solution of
an uncharged dilatonic black hole.

We can choose a gauge in which the Higgs field $\Phi$ and the gauge field
$A_{\mu}$ have the form
\bsq
\bq
&&\Phi = \eta X(r,\theta) e^{i\varphi}
\label{5higgscampo}\\
&&A_{\mu} = \frac{1}{e} \left(P(r,\theta \right)-1)\delta_{\mu}^{\varphi}
\label{5gaugecampo}
\eq
\esq
i.e.\ we are considering a winding number $1$ vortex.
Substituting these forms into the vortex equations of motion
we obtain
\bsq
\bq
&&-\frac{1}{r^2} \left[r \left(r-2E \right)X_{,r} \right]_{,r}
-\frac{1}{r^2 \sin\theta} \left[\sin\theta \, X_{,\theta} \right]_{,\theta}
+ \frac{X}{2} (X^2 - 1)
+ \frac{XP^2}{r^2 \sin^2\theta} = 0
\label{5xequation}\\
&&\left [ \left (1-\frac{2E}{r} \right ) P_{,r} \right]_{,r}
-\frac{X^2 P}{\beta}
+ \frac{\sin\theta}{r^2} \left [\frac{P_{,\theta}}{\sin\theta}
\right ]_{,\theta} = 0.
\label{5pequation}
\eq
\esq

To argue the existence of a vortex solution analytically, we assume that
the black hole is large compared to the string width, i.e.\ $E \gg 1$.
We then take $X = X(R)$, $P = P(R)$ with $R = r \, \sin\theta$,
and substituting in the vortex equations above, 
denoting the derivative with respect to $R$
by a prime,  we get
\bsq
\bq
\left [-1+\frac{2E}{r} \sin^2\theta \right] \left [ X'' +\frac{X'}{R} \right]
+\frac{X}{2} \left (X^2-1 \right) + \frac{X P^2}{R^2} &=& 0 \\
\left[1-\frac{2E}{r} \sin^2\theta \right] \left [ P'' -\frac{P'}{R} \right] 
-\frac{X^2 P}{\beta} &=& 0.  
\eq
\esq
These can be seen to be the Nielsen-Olesen equations, (\ref{basic}),
up to terms of the form $ \frac{2E}{r} \sin^2\theta $ times derivatives of
$X$ and $P$.  In and near the core, where $R = r\,\sin\theta \leq 1$, 
$\sin\theta= {\cal O} (\frac{1}{r}) \leq {\cal O} (\frac{1}{E})$; so
in this thin vortex limit, these corrections
are negligible, and therefore to a good approximation the 
vortex equations are identical to the Nielsen-Olesen ones (\ref{basic}),
and the Nielsen-Olesen solution is still a good solution
in and near the core of a thin vortex even
at the event horizon (as proven in \cite{AGK} using Kruskal coordinates)
and the string simply continues regardless of the black hole
as confirmed numerically in \cite{AGK}.

We now generalise these results to 
charged black holes in the presence of a dilaton.
These are solutions to the equations of motion of the low energy effective 
action in the Einstein frame \cite{RJ,HH,GHS}
\be
S_{dil-Max} = \int  d^4 x \sqrt{-g} \left [ 
- R + 2 (\nabla\phi)^2 - V(\phi)
- e^{-2 \phi} F^2
\right ]
\ee
where $F$ is electromagnetic field strength of the Maxwell field 
which does not interact directly with the Higgs field, and we will take
$V(\phi) = 2 M^2 \phi^2$ to be the dilaton potential for the purposes
of this paper. In general the dilaton potential could be more
complicated, but this quadratic approximation should be valid while
the dilaton remains close to its minimum. 

A general spherically symmetric black hole solution has a metric
of the form
\be\label{sphsym}
ds^2 = \lambda(r) \, dt^2
- \frac{1}{\lambda(r)} \, dr^2
- C^2 \,  (d\theta^2
+ \sin^{2}\theta \, d\varphi^2)
\ee
in which the electromagnetic equation of motion has 
the general magnetic solution
\be\label{magnet}
F = Q \sin\theta d\theta \wedge d\phi 
\ee
and the equations of motion in the Einstein frame are \cite{RJ,HH}
\bsq\label{masseqns}
\bq
\left[C^2 \, \lambda \, \phi^\prime \right]^\prime
&=& M^2 C^2 \phi - \frac{Q^2}{C^2}\, e^{-2\phi} \label{massd}\\
\left[C^2 \, \lambda^\prime \right]^\prime
&=& -2M^2 C^2 \phi^2 + \frac{2Q^2}{C^2}\, e^{-2\phi} \label{masslam}\\
\left[\lambda \,\left(C^2 \right)^\prime \right]^\prime
&=& 2 - 2M^2 C^2 \phi^2 -\frac{2Q^2}{C^2}\, e^{-2\phi}
\label{massc}\\
0 &=& C^{\prime\prime}(r) + C \, \phi^{\prime 2}
\label{massc2}
\eq
\esq
The electric solution is obtained by applying an electromagnetic 
duality transformation to the equations of motion that preserves the metric
but changes the sign of the dilaton and is explicitly given by 
\bsq\label{emdual}
\bq
\phi_E &=&  - \phi_M\\
F_{ab}^E &=& \frac{1}{2} \, e^{-2 \phi_M}  
\, \epsilon_{ab}\,^{cd} \, F_{cd}^M
\eq
\esq

In general (i.e.\ for nonvanishing dilaton potential) the solutions to
these cannot be expressed in closed
analytic form, however, for the moment proceeding as in 
\cite{AGK} and \cite{BRG} we look for a vortex 
solution by taking
$ X = X(\sigma), P=P(\sigma)$, with $\sigma = Ce^{\phi}\sin\theta$
which gives the vortex equations
\bsq
\bq
&&\frac{{\dot X}}{\sigma} \, \left[-1 + \sin^{2}\theta \,
\left(2 - \lambda \, C^2 \, \frac{(Ce^{\phi})^{\prime\prime}}
{Ce^{\phi}} - C^2 \, \frac{(Ce^{\phi})^\prime}
{Ce^{\phi}} \, \left[2\lambda\, \frac{(Ce^{\phi})^\prime}
{Ce^{\phi}} + \lambda^\prime + 2 \, a \, \lambda \, \phi^\prime
\right]\right)\right]
\nonumber\\ &&\,\,\,\,\,\,
+{\ddot X} \, \left[-1 + \sin^{2}\theta \,
\left(1 - \lambda \, C^2 \, \left[\frac{(Ce^{\phi})^\prime}
{Ce^{\phi}}\right]^2\right) \right] +
\frac{XP^2}{\sigma^2}+\frac{X}{2}
(X^2 - 1) = 0
\label{5vortttxmass}\\
&&\frac{{\dot P}}{\sigma} \, \left[-1 + \sin^{2}\theta \,
\left(\lambda^\prime \, C^2 \, \frac{(Ce^{\phi})^\prime}
{Ce^{\phi}} + \lambda \, C^2 \,
\frac{(Ce^{\phi})^{\prime\prime}}{Ce^{\phi}}
+ 2 \, a \, \lambda \, \phi^\prime \, C^2 \,
\frac{(Ce^{\phi})^\prime} {Ce^{\phi}}\right) \right]
\nonumber\\ &&\,\,\,\,\,\,
+{\ddot P} \, \left[1-\sin^{2}\theta \,
\left(1 - \lambda \, C^2 \, \left[\frac{(Ce^{\phi})^\prime}
{Ce^{\phi}}\right]^2\right) \right] - \frac{X^2 P}{\beta} = 0
\label{5vortttpmass}
\eq
\esq
where a dot means the derivative with respect to $\sigma$.
These equations (\ref{5vortttxmass})-(\ref{5vortttpmass})
are the Nielsen-Olesen ones up to terms
which may be written as
\bsq\label{corrterms}\bq
{\cal T}_1 &=& {\sigma^2\over C^2 e^{2\phi}} \left ( 1 - \lambda \, C^2
\, \left[\frac{(Ce^{\phi})^\prime} {Ce^{\phi}}\right]^2 \right )\\
{\cal T}_2 &=& {\sigma^2\over C^2 e^{2\phi}} \left (
\lambda^\prime \, C^2 \, \frac{(Ce^{\phi})^\prime}
{Ce^{\phi}} + \lambda \, C^2 \,
\frac{(Ce^{\phi})^{\prime\prime}}{Ce^{\phi}}
+ 2 \, a \, \lambda \, \phi^\prime \, C^2 \,
\frac{(Ce^{\phi})^\prime} {Ce^{\phi}} \right )
\eq\esq
multiplied by derivatives of the vortex fields.
Provided these correction terms are negligible
in and near the core of a thin
vortex, the Nielsen-Olesen solutions will be a good approximation
to the string threading the black hole.
 
Having derived the general equations, we now look at electrically
and magnetically charged black holes with a massless and massive
dilaton in turn.

\subsection{\bf Charged black holes with massless dilaton.}

When the dilaton is massless ($M = 0$), the black hole
solution of the equations (\ref{masseqns})
with a pure magnetic charge $Q$ is \cite{GHS}
\bsq
\bq
ds^2 &=& \left(1-\frac{2E}{r} \right) dt^2 -
\left(1-\frac{2E}{r} \right)^{-1} dr^2
- r \left(r - \frac{Q^2}{E} \right) ( d\theta^2
+ \sin^{2}\theta \, d\varphi^2 )
\label{mslsmet}\\
e^{-2\phi} &=& 1 - \frac{Q^2}{Er}
\label{5C}
\eq
\esq
the mass, $E$, and the charge, $Q$, are written in ``vortex units'', and
are related by $Q^2 \leq 2 E^2$.

We may therefore read off
\bsq\bq
{\cal T}_1 &=& {\sigma^2\over r^2} {2E\over r}\left [ 
1 + {Q^2 \over 2E^2} - {Q^2 \over Er} \right ] \label{t1mag} \\
{\cal T}_2 &=& {\sigma^2\over r^2} \left [ {2E\over r}
\left (1 - {Q^2 \over Er} \right ) - {aQ^2 \over Er} 
\left ( 1 - {2E\over r} \right ) \right ] \label{t2mag}
\eq\esq

In and near the core of a thin vortex the charged correcting 
terms like (\ref{t1mag}) are always negligible 
when compared with 
the Nielsen-Olesen ones, as they are of order 
${\cal O} (\frac{1}{E^2})$, 
while the dilatonic coupling ones like the second part of 
(\ref{t2mag}) are of order
${\cal O} (\frac{a\,Q^2}{E^4})$
and therefore could be relevant for extremely large couplings of the
dilaton to the vortex $|a| \geq {\cal O} (\frac{E^4}{Q^2})
\geq {\cal O}(E^2)$, however, these are not particularly 
realistic values (e.g. $|a| = 0, 1, \sqrt3$ is usual).
Therefore to a good approximation the vortex solution is given by
the Nielsen-Olesen solution, and since $\sigma = r\sin\theta$ for the
magnetic black hole, the solution is in fact identical to the Schwarzschild
vortex.

We also note that these conclusions do not change with $\frac{Q}{E}$ and so 
still apply in the particular case where the black hole is extremal 
$Q^2 = 2 E^2$.  In this case the horizon is singular in the Einstein frame
with a vanishing area \cite{GHS}, however in the string frame
\be
ds^2 = dt^2 - \left(1-\frac{2E}{r}\right)^{-2} dr^2
- r^2 (d\theta^2
+ \sin^{2}\theta \, d\varphi^2 ),
\ee
and the previously singular horizon $r = 2E$
has been pushed off to infinite proper distance. Whether one could say that
the string was or was not piercing the horizon is a 
moot point.

Let us now consider a pure electrically charged
black hole with a massless dilaton, given by the duality transformation
(\ref{emdual}), which has the same metric 
as the magnetic black hole, but the dilaton is now given by
\be\label{5dilect}
e^{2\phi} = 1 - \frac{Q^2}{Er}
\ee

Now we obtain
\bsq\bq
{\cal T}_1 &=& {\sigma^2\over(r - Q^2/E)^2} {(2E - Q^2/E)\over (r - Q^2/E)}\\
{\cal T}_2 &=& {\sigma^2\over(r - Q^2/E)^2} \left [
{2E\over r} + {Q^2 a\over Er} {(r-2E)\over (r-Q^2/E)} \right ]
\eq\esq
Clearly, when $Q^2 < E^2$ these terms are negligible for similar 
reasons as before.
However, consider now the extremal (or near extremal) case
$Q^2 = 2 E^2 - qE$. In this case, we see that 
\bsq\bq
{\cal T}_1 &=& {\sigma^2 q \over (r-2E+q)^3} < O(\sigma^2/q^2)\\
{\cal T}_2 &=& {\sigma^2\over (r-2E+q)^3} {2E\over r}
\left [ q + (r-2E)(1+a - {aq\over 2E}) \right ]< O(\sigma^2/q^2)
\eq\esq
We now see that close to the extremal limit, the Nielsen-Olesen
approximation breaks down in the vicinity of
the horizon. What this means is that the thin vortex limit 
has broken down, and our analytic approximation
is no longer valid. However, if we examine the area of the horizon,
$4\pi C^2 = 8\pi Eq$, we see that we might only reasonably expect a thin
vortex approximation to work for $Eq \gg 1$, (or $q\gg1$ if we look at
the string frame), therefore, the breakdown of this method is due to the
breakdown of the coordinate system at the horizon, which becomes singular
in the extremal limit.

At extremality, ${\cal T}_1=0$,
and ${\cal T}_2 = {2E\sigma^2 (1+a) \over r(r-2E)^2}$.
For $a \not= -1$ these terms eventually become important in the core when
$(r - 2E)^2 \, r \leq 2 E$
i.e.\ close to the horizon (which is also singular).
As the size of the black hole is now zero
this means that in fact the string,
instead of penetrating the black hole, swallows it. Again 
this result does not depend on the frame. 
Note however, that for $a = -1$, our analytic approximation is
exact and the Nielsen-Olesen solution gives the form of the
string. Since $\sigma=0$ on the horizon, one could say that the
flux of the string was expelled.

\subsection{\bf Charged black holes with massive dilatons.}

When the dilaton is massive the character of the black hole 
background is in general different from the massless one as 
(\ref{mslsmet}), (\ref{5C}) and (\ref{5dilect})
are no longer solutions of the geometry equations (\ref{masseqns}).
Qualitatively speaking there are three distinct types of black hole 
\cite{RJ,HH}, depending on the relative sizes of the black hole, $E$,
and the Compton wavelength of the dilaton, $M^{-1}$. 
Black holes which are small compared to the Compton wavelength of the
dilaton ($EM\ll1$) resemble the massless dilaton solutions
already discussed, which have the causal structure of a Schwarzschild
black hole -- a single horizon and spacelike singularity.
Those black holes which are large compared to the Compton wavelength of the 
dilaton ($EM \gg 1$) resemble the Reissner-Nordstr\o m solution in the
region exterior to the horizon, although it is possible that their overall
causal structure is quite different in that there can be one, two or even
three horizons \cite{RJ,HH}. The intermediate case $EM = O(1)$, is the
borderline between these two behaviours, where additional horizons are
possible and even a special extremal solution with a triply degenerate
horizon occurs. These black holes have no approximate analytic description.

When the Schwarzschild radius $E$ is less than the
Compton wavelength of the dilaton, i.e.
$E \ll M^{-1}$, the black hole does not see
the mass of the dilaton and behaves like the massless case,
and therefore (\ref{mslsmet}),(\ref{5C}) and (\ref{5dilect})
are good approximation to the true black hole background solution.
We therefore expect the results of the previous subsection to apply,
and in the thin vortex limit the vortex will be given by the Nielsen-Olesen
solution. Since $1\ll E \ll M^{-1}$ the dilaton is also effectively massless
as far as the string is concerned.
(Although for a minimal dilaton mass of $m = 10^3 Gev$
this means black hole masses of rather less than $10^{11}$g,
and hence would require a primordial black hole.)

When the Schwarzschild radius $E$ is much larger than the   
Compton wavelength of the dilaton $\frac{1}{M}$, i.e.
$E \gg \frac{1}{M}$, the dilaton (and corrections to the geometry) are
of order ${Q^2\over M^2r^4} \leq O({1\over M^2E^2})$
and hence we can regard the dilaton as being essentially fixed
and the geometry as the Reissner-Nordstr\o m one 
\be
ds^2 = \left(1-\frac{2E}{r}+\frac{Q^2}{r^2}\right) \, dt^2
- \left(1-\frac{2E}{r}+\frac{Q^2}{r^2}\right)^{-1} \, dr^2
- r^2 ( d\theta^2
+ \sin^{2}\theta \, d\varphi^2 )
\ee
which is now being extremal for $|Q| \simeq E$.
We can now use the results of \cite{BG,BRG} to conclude that in the thin
vortex limit, the Nielsen-Olesen solution is a good approximation to
the vortex, and for extremal black holes there is a flux expulsion phenomenon
when the thickness of the string core becomes comparable to the black 
hole horizon scale.

We now consider black holes for which Schwarzschild radius 
is similar in scale to the Compton wavelength of the 
dilaton $EM\simeq 1$. In this case, there is no simple analytic
form for the geometry, and we must estimate the correcting terms
(\ref{corrterms}) from the equations of motion. We first note that if the
charge of the black hole is small, the dilaton field will not differ
much from its vacuum value, being of order $Q^2/E^2$ for $EM\simeq 1$.
Therefore the interesting r\'egime in which to analyse the vortex is
close to the extremal limit, $QM = O(1)$. One of the interesting 
features of massive dilatonic black holes is that they possess a
richer horizon structure than that of the massless dilatonic solutions.
In particular, at $QM = e/2$, there is a phase transition in the 
types of extremal solutions possible. For $QM<e/2$ there is only one
horizon, and the extremal solution corresponds to the singularity moving
out to the black hole horizon (i.e.\ $C=0$). For $QM>e/2$ an
extremal solution similar to the Reissner-Nordstr\o m one occurs, in that
$\lambda = \lambda' = 0$ at the horizon. For $QM=e/2$, there is a special
triply degenerate extremal solution, where $\lambda, \lambda'$ and 
$\lambda''$ all vanish. For all values of $QM$ however, the solutions do have
the common feature that $\phi$ is decreasing (increasing) outside the
horizon for the magnetic (electric) black hole, that $\lambda$ monotonically
increases from 0 to 1 outside the horizon, and finally that $C'\geq 1$ 
outside the horizon. We therefore need to estimate the ${\cal T}_i$
with all this in mind.

First note that
\be
(C'+c\phi')^2 \leq | C^{\prime2} - C\phi^{\prime2} |
= \left | 1 - \lambda CC' - M^2C^2\phi^2 - {Q^2 \over C^2 e^{2\phi}} \right |
\ee
using (\ref{massc},\ref{massc2}), hence
\be
{\sigma^2\over C^2e^{2\phi}} \geq {\cal T}_1 \geq
-{\sigma^2\over C^2e^{2\phi}}  \left | \lambda CC' + M^2 C^2 \phi^2 +
{Q^2 \over C^2 e^{2\phi}} - 1 \right |
\ee
Then, using (\ref{massd},\ref{massc}) one can show
\be
{\cal T}_2 = {\cal T}_1 + {\sigma^2\over C^2e^{2\phi}} 
\left [ M^2C^2 \phi (1 - \phi) - 2{Q^2 \over C^2e^{2\phi}}
+2(a+1)\lambda \phi' (C'+C\phi')\right ]
\ee
(For the electric black hole, replace $\phi$ by $|\phi|$ except in the 
initial ${\sigma^2\over C^2e^{2\phi}} $ term.)
Since these terms, for reasonable $a$, and with the possible exception of 
${Q^2 \over C^2e^{2\phi}}$, can be shown to be at most of order unity, the
magnitude of the ${\cal T}_i$ boils down to the minimal value of
$Ce^\phi$. In most cases, this quantity attains its minimum on
the horizon, however, for a small subset of solutions (namely those
close to extremal, for which the value of the dilaton on the horizon,
$\phi_h$, lies approximately in the range $[1-1/\sqrt{2},1]$), $Ce^\phi$
actually has its minimum outside the horizon, and the spacetime in the
string frame has a wormhole-like structure \cite{HH}.

We begin therefore by estimating $Ce^\phi$ on the horizon.
Starting with the magnetic black hole and evaluating (\ref{massd},\ref{massc})
at the horizon, using the properties of the dilaton and metric functions,
one can readily obtain the following inequalities for $C_h$:
\bsq\label{cineq}\bq
C_h^2 &\leq& {Qe^{-\phi_h} \over M\sqrt{\phi_h}} \\
{1\over 2 M^2 \phi_h^2} \left [ 1 - \sqrt{1 - 4M^2Q^2 \phi_h^2 e^{-2\phi_h}}
\right ] \leq &C_h^2 & \leq {1\over 2 M^2 \phi_h^2} \left [ 1 +
\sqrt{1 - 4M^2Q^2 \phi_h^2 e^{-2\phi_h}}\right ] 
\eq\esq
Hence
\be
C_h^2 e^{2\phi_h} \geq {Q^2\over 2M^2Q^2 \phi_h^2 e^{-2\phi_h}} \left [
1 - \sqrt{1 - 4M^2Q^2 \phi_h^2 e^{-2\phi_h}}\right ] 
\geq Q^2
\ee
If $Ce^\phi$ is minimised at the horizon, then clearly ${\cal T}_i = 
O(Q^{-2}) = O(E^{-2})$, and the thin vortex approximation is satisfied.
If $Ce^\phi$ is not minimised at the horizon, then we note that
the value of $\phi$ required is of order unity \cite{HH}, hence
O($Ce^\phi|_{\rm min}) >$ O($C_h$) $\simeq$ O($Q$), and so ${\cal T}_i = 
$O$(E^{-2})$ in this case as well. Therefore magnetic black holes always
admit a thin vortex approximation.

For the electric black hole, the inequalities (\ref{cineq}) are still
valid, provided we replace $\phi_h$ by $|\phi_h|$. This however means
that $C_h^2 e^{2\phi_h} = C_h^2 e^{-2|\phi_h|}$, and hence
\bq
{e^{-2|\phi_h|} \over 2M^2 \phi_h^2} \left [\textstyle{1 -  \sqrt{1 - 4M^2Q^2 
\phi_h^2 e^{-2|\phi_h|}}}\right ] 
&\leq& C_h^2 e^{-2|\phi_h|} \nonumber\\
&\leq& {\rm Min} \left \{
{Qe^{-3|\phi_h|} \over M\sqrt{|\phi_h|}}, \;\;\;
{e^{-2|\phi_h|} \over 2M^2\phi_h^2} \left [ \textstyle{1 + \sqrt{1 - 4M^2Q^2 
\phi_h^2 e^{-2|\phi_h|}}}\, \right ] 
\right \}
\eq
which gives no satisfactory bound on $C_h^2e^{2\phi_h}$, as might have been
expected, given the massless electric black hole. We therefore suspect
that electric black holes are closer to their massless counterparts, in
that unless $QM>e/2$, (so that $|\phi_h|<1$), 
nearly extremal electric black holes will have no analytic thin
vortex approximation for the vortex.

\subsection{Extremal black holes and flux expulsion}

We would now like to make some comments about whether the phenomenon 
of flux expulsion can occur in the extremal limit. Note first that for
the massless extremal electric black hole (and for $MQ<e/2$), the vortex
is trivially expelled from the black hole, since the area of the black hole
is zero in both the Einstein and string frames. For the magnetic black holes
(and for $QM>e/2$ electric black holes) we must however look more closely
at the system. For the extremal black hole, the equations for the vortex
on the horizon reduce to
\bsq\bq
-{1\over \sin\theta} \partial_\theta \left ( \sin\theta \partial_\theta
X \right ) + {XN^2P^2\over \sin^2\theta} + {\half} C_h^2 e^{2\phi_h} 
X(X^2-1) &=& 0 \label{horxeq} \\
\sin\theta\partial_\theta \left [{\partial_\theta P \over \sin \theta} \right
] - {C_h^2 e^{2\phi_h}X^2P\over \beta} &=& 0, \label{horpeq}
\eq\esq
for all values of dilaton mass for magnetically charged black holes, and
for $QM>e/2$ if electrically charged.
These equations are identical in form to the Reissner-Nordstr\o m equations
of \cite{BRG}, and we can therefore simply use the result from that paper
that the vortex flux lines {\it must} be expelled from the extremal black hole
if
\be
{{\cal M}^5 \over (1-{\cal M})^2} < {3\sqrt{3} \over 2\pi^2} {\beta^2\over
N^4} \simeq {\beta^2\over 4N^4}.
\ee
where ${\cal M} = C_h e^{\phi_h} / \sqrt{2}$, $\beta$ is the Bogomolnyi
parameter, and we have reinstated $N$, the winding number introduced 
in section~\ref{dilsec}.  For $N=\beta=1$, this gives the (weak) 
bound that for $C_h e^{\phi_h} < 0.7$, flux expulsion must occur.

For the massive dilatonic extremal black holes with $QM>e/2$, we are
able to evaluate $C_h$ and $\phi_h$ exactly from the equations of motion
at the horizon \cite{RJ} giving
\bsq\bq
|\phi_h| (|\phi_h|+1)^2 e^{-2|\phi_h|} &=& {1\over M^2Q^2} \label{phiex}\\
C_h^2 &=& {Q\over M} |\phi_h|^{-1/2} e^{-|\phi_h|}\label{cex}
\eq\esq
where consistency with (\ref{masslam}) requires that the $\phi_h\leq 1$ root
of (\ref{phiex}) be taken.

For the magnetic black holes we therefore obtain
\be
0.7 > Q (\phi_h+1)^{1/2} > {e\over 2M}
\ee
hence $Q<0.7$, $M>1.94$ are minimal requirements for flux expulsion to occur. 
For the electric black hole with $QM>e/2$:
\be
0.7 > Q(|\phi_h|+1)^{1/2} e^{-2|\phi_h|} > {\sqrt{2}Q\over e^2} > 
{1\over \sqrt{2}eM}
\ee
giving $Q<3.65$, $M>0.37$ as minimal requirements for flux expulsion.
For the massless extremal magnetic black hole, $C_he^{\phi_h} = 2E$,
hence $E<0.35$ is the appropriate bound  in this case.

To sum up, in this section we have shown that in a wide variety of cases,
the Nielsen-Olesen solution gives a good approximation to the thin vortex
solution in the presence of a black hole. 
The only situation in which it does not give
an adequate description is that of near extremal electric black holes
where the dilaton is either massless, or has a low mass. In this case,
a full numerical study would be required\footnote{Such a study has been
performed by Moderski and Rogatko\cite{MRpc} for a particular `$a$', 
and we understand that they do indeed observe the flux expulsion phenomenon.}.
Since none of these arguments rest on the fact that the string must
thread the black hole, we may conclude, as in \cite{AGK},
that these arguments can be used to construct strings terminating on
black holes. Finally, we have demonstrated that we can prove the 
expulsion of flux from a range extremal black holes with sufficiently small
Schwarzschild radii.

\section{\bf Gravitating strings.}\label{gravsec}

In this section we consider the gravitational back-reaction of a 
thin vortex on the spacetime geometry and dilaton, using the same 
method as in \cite{AGK,GH,BRG}, i.e., expanding the equation of motion
in powers of $\epsilon$, the gravitational strength of the string, which
is assumed small. 
Before starting, it is worth asking what sort of solutions we expect 
to obtain; for the Einstein string the known asymptotic metrics
were the AFV, Israel-Kahn, and C-metrics. The Israel-Khan
metric, which is uncharged, will also be a solution in dilatonic gravity.
The dilatonic C-metrics were given by Dowker et.\ al.\ in \cite{DGKT},
these metrics are the generalisation of the
C-metrics of Kinnersley and Walker \cite{KW}, and consist of
a black hole under constant acceleration, driven by a conical singularity
extending from the event to the acceleration horizon. Based on this, one 
expects that the generalisation of the AFV metric is the dilatonic black hole
metric (\ref{sphsym}), with either a massive or massless dilaton, with a
conical slice removed.  As we will see, because of our choice of the
arbitrary coupling parameter $a$, the actual set of solutions obtained
is somewhat more complex.

We begin by considering the most general static axially symmetric metric 
\be\label{aximet}
ds^2 = e^{2\psi} dt^2 
- e^{2(\gamma-\psi)}\left(d\zeta^2 + d\rho^2\right)
-\alpha^2 e^{-2\psi} \, d\varphi^2
\ee
where $\psi$, $\gamma$, $\alpha$ are functions of $\zeta$ and $\rho$
and the coordinates are given in ``vortex units''. In these coordinates,
the function appearing in the analytic thin vortex approximation is
now $\sigma = \alpha e^{\phi-\psi}$. In order for this approximation to 
hold, the equations of motion for $X$ and $P$ imply that 
\bsq\bq
\sigma_{_{,}i}^2 &=& e^{2\phi} e^{2(\gamma - \psi)} + O(E^{-1}) \\
\sigma_{_{,}ii} + {\sigma_{_{,}i} \alpha_{_{,}i}\over\alpha}
+ 2(a+1) \sigma_{_{,}i} \phi_{_{,}i} &=& {e^{2\phi} e^{2(\gamma - \psi)} 
\over \sigma} + O(E^{-1}) 
\eq\esq
throughout the core of the string. Applying this to the energy-momentum 
tensor for the vortex from  (\ref{stringem}), gives
\bsq
\bq
{\hat T}^t_t &=& e^{(4+2a)\phi} \left [ {X^2P^{2}\over \sigma^{2}} + {\quarter} 
(X^{2}-1)^{2} + e^{-2(\gamma+\phi-\psi)}   \left( X_{,i}^2 
+ \frac{\beta}{\sigma^2} P_{,i}^2  \right ) \right ] 
\simeq e^{(4+2a)\phi}{\cal E}_{_0}(\sigma) \label{temom}\\
{\hat T}^\varphi_\varphi &\simeq& -e^{(4+2a)\phi} {\cal P}_{_0\varphi}(\sigma)
\label{phiemom}\\
{\hat T}^\zeta_\zeta &+& {\hat T}^\rho_\rho 
\simeq  e^{(4+2a)\phi} [
{\cal E}_{0}(\sigma) - {\cal P}_{_0R}(\sigma)] \label{tremom}
\eq
\esq
where we have used the notation of (\ref{NOtens}) 
In these coordinates the (relevant) equations of motion are:
\bsq
\bq
\alpha_{,ii} &=& - \sqrt{-g} \left[ 2M^2\phi^2 + \epsilon e^{(4+2a)\phi} 
\left ( {\cal E}_{0}(\sigma) - {\cal P}_{_0R}(\sigma) 
\right) \right ] \label{alphaeq}\\
\left(\alpha \psi_{,i} \right)_{,i}  &=& -{\half} \sqrt{-g}
\left[2M^2\phi^2 - e^{-2\phi} | F^2| - \epsilon e^{(4+2a)\phi}
\left ( {\cal P}_{_0R}(\sigma) + {\cal P}_{_0\varphi}(\sigma) \right ) 
\right] \label{psieq}\\
\gamma_{,ii} &=&  - \psi_{,i}^2 - \phi_{,i}^2 -\frac{\sqrt{-g}}{\alpha}
\left[ M^2 \phi^2 - {\half} e^{-2\phi} |F^2| - \epsilon e^{(4+2a)\phi} 
{\cal P}_{_0\varphi}(\sigma) \right ] \label{gammaq}\\
( \alpha \phi_{,i} )_{,i} &=&  \sqrt{-g} \left [ M^2 \phi 
+ {\half} e^{-2 \phi} {F}^2 - \epsilon e^{(4+2a)\phi} \left( {\half} 
\left ( {\cal P}_{_0R}(\sigma) + {\cal P}_{_0\varphi}(\sigma) \right ) 
+ (1+a) {\cal E}_{0}(\sigma) \right ) \right ] \label{dileq}\\
0 &=&\left[e^{-2\phi} \alpha F_i^{\ \mu} \right ] _{,i} \label{maxeq}
\eq\esq
(where $i = \rho,\zeta$, and the summation convention applies).
Hence we see that the source terms in the Einstein equations 
consist of terms which are functions of the original spherical 
$r$-coordinate, and the vortex function, $\sigma$.

For example,  the massless dilaton black hole in axisymmetric coordinates is
\bsq
\bq
\alpha_0 &=& \rho \label{alphaback}\\
e^{2\psi_0} &=& {R_+ +R_- -2\Delta\over R_+ +R_- +4E-2\Delta} \label{psiback}\\
e^{2\gamma_0} &=& {\left(R_{+}+R_{-} \right)^2 -4\Delta^2\over4R_{+}R_{-}}
\label{gammaback}\\
e^{\pm2\phi_0} &=& {R_+ + R_- + 4E + 2 \Delta 
\over R_+ + R_- + 4E - 2\Delta} \label{dilback}
\eq
\esq
where
\be
R_{\pm}^2=\rho^2+ \left[\zeta\pm \Delta\right]^2 =\left [ r-2E
+\Delta\pm\Delta \cos\theta\right ]^2
\ee
with $\Delta = E -\frac{Q^2}{2E}$. This is obtained by 
using the coordinate transformation
\bsq
\bq
\zeta &=& \left(r - E - \frac{Q^2}{2E}\right)\cos\theta
\label{zetadefin}\\
\rho^2 &=& \left(r - \frac{Q^2}{E}
\right)(r - 2E)\sin^2\theta
\label{rodefin}\
\eq
\esq

We begin by examining the effect of the dilatonic vortex threading the 
Schwarzschild black hole since the lack of electromagnetic charge
considerably simplifies the problem.  First note that to O$(\epsilon)$
the geometry is unaffected by the dilaton, and only reacts to the
vortex energy-momentum. The metric is therefore given by
the results in \cite{AGK}, giving
\be
ds^2 = \left(1-\frac{2{\tilde E}}{{\tilde r}_s} \right) d{\tilde t}^2
- \left(1-\frac{2{\tilde E}}{{\tilde r}_s}\right)^{-1} d{\tilde r}_s^2
- {\tilde r}_s^2 \, d\theta^2
- {\tilde r}_s^2 \, (1 -\epsilon A)^2 \, e^{-2\epsilon D} \, \sin^{2}\theta \,
d\varphi^2
\ee
where the time, $t$, has been rescaled to the proper time at asymptotic
infinity, ${\tilde t} = e^{\frac{D}{2}} \, t$, etc. This metric is clearly
that of a  Schwarzschild black hole with renormalised mass  
${\tilde E} = e^{\frac{\epsilon D}{2}} \, E$, 
with a deficit angle of $2\pi \epsilon(A+D) = \epsilon\mu$
(independent of the radial stresses),
and an apparent conical singularity which is of course smoothed out
by the vortex.  When the radial stresses do not vanish 
($\beta \not= 1$) there is a red/blue-shift
of time between infinity and the core of the string \cite{AGK}.

We now calculate up to O$(\epsilon)$,
the back reaction of the vortex
on the dilaton. We use the spherical, Schwarzschild, coordinates
for simplicity. Assuming a form $\phi = \epsilon f_s(R)$ where $R=r\sin\theta$
and $f_s$ is the pure dilatonic cosmic string solution given in 
\cite{GS} and reviewed in section 2, we obtain
\bq  
\left(1 - \frac{2E}{r} \sin^2\theta \right)
\left[f_s^{\prime \prime} + \frac{f_s^{\prime}}{R}
\right]
&-& M^2 f_s + {\half} ({\cal P}_{_0R}(R) + {\cal P}_{_0\varphi}(R))
- (1+a) {\cal E}_{_0}(R) \nonumber \\
&=& - \frac{2E}{r^3} \sigma^2 
\left[f_s^{\prime \prime} + \frac{f_s^{\prime}}{R} \right] = 0
\eq
For $M^2=0$, this equation is clearly satisfied to order O($E^{-2}$)
since the dilaton is logarithmic outside the core.
For $M^2 \neq 0$, the situation is slightly more subtle. If the Compton
wavelength of the dilaton is much greater, or much less, than the
Schwarzschild radius, then the equation is valid, since the dilaton
will either be qualitatively massless, or at its vacuum value near the
horizon. However, for $M^{-1}\simeq E$, this analytic approximation will
not hold, and the functional form of the dilaton will be modified
in the vicinity of the horizon. In all cases however, this approximation
holds for large radius.
 
This shows that the vortex switches
on a non-vanishing dilaton field on the horizon of the black hole,
$\phi = \epsilon f_s(2E\sin\theta)$, which means that there is an
effective dilatonic charge for the massless dilaton 
of ${\cal D}_1 = 2E(a+1) \epsilon {\hat \mu}$, in other words, the charge 
generated by a fragment of cosmic string of length $2E$. In this
sense, the system behaves very much as if it can ``see'' the fragment
of string behind the event horizon.

Moving to the charged black holes, first note that the existence of a 
dilatonic vortex breaks the electromagnetic duality invariance via the
presence of the ${\cal E}_0$ etc.~terms in (\ref{dileq}) which only vanish
for $\beta = -a = 1$.
We will therefore have to consider electric and magnetic black holes
seperately.  To zeroth order we have the background solutions 
(\ref{alphaback})-(\ref{gammaback}) and using  \cite{BRG} as a guide, we guess
that the perturbed solution takes the form:
\bsq\bq
\alpha &=& \alpha_0 \left ( 1 + \epsilon e^{2(a+1)\phi_0} b(\sigma)\right)\\
\psi &=& \psi_0 + \epsilon e^{2(a+1)\phi_0} d(\sigma) \\
\gamma &=& \gamma_0 +  \epsilon e^{2(a+1)\phi_0}  g(\sigma) \\
\phi &=& \phi_0 + \epsilon e^{2(a+1)\phi_0} f(\sigma) \\
A_\mu &=& A_{0\mu} \left ( 1 + \epsilon e^{2(a+1)\phi_0}  q(\sigma)\right )
\eq\esq

Inputting these into the equation of motion gives, after some algebra,
and to order O($E^{-2}$):
\bsq\bq
b'' + {2b'\over\sigma} &=& - [ {\cal E}_{_0} - {\cal P}_{_0R} ] \\
d'' + {d\over \sigma} &=& {\half} [ {\cal P}_{_0R} + {\cal P}_{_0\varphi}]\\
g'' &=& {\cal P}_{_0\varphi} \\
f'' + {f'\over\sigma} &=& M^2 f + (a+1){\cal E}_{_0} - {\half} \left (
{\cal P}_{_0R} + {\cal P}_{_0\varphi} \right ) \\
(\sigma^3 q_M')' &=& 2\sigma^2 \left ( b' + 2f'-2d' \right )\\
q_e'' + {q_E'\over\sigma} &=&0 
\eq\esq
where the subscripts $M$ and $E$ indicate the magnetic and electric 
corrections respectively.
Note that these equations are valid only in the vicinity of the core, and
only to O($E^{-2}$), outside the core, where the terms no longer involve
the vortex core, and are typically of order $E^2/r^4$, the equations differ
depending on whether the dilaton is massive or massless, and whether the
black hole is electrically or magnetically charged.

These equations are readily integrated to obtain for the leading order 
correction
\bsq\bq
b &=& - A(\sigma) + {B(\sigma)\over\sigma} +b_0\\
d &=& {\half} D(\sigma) + d_0 \\
g &=& D(\sigma) +  g_0 \\
f &=& f_s(\sigma) + f_0 =_{_{M=0}} - {\half} D(\sigma)+ f_0 
+ (a+1) \int_0^\sigma {A(\sigma) + D(\sigma) \over\sigma}\\
q_M &=& b+2(f-d) +  q_{0M}+ {1\over \sigma^2} \int_0^\sigma \left [ B(\sigma)
+ 2\sigma (a+1) \left (A(\sigma) + D(\sigma) \right ) \right ] \\
q_E &=& q_{0E}
\eq\esq
where the integration constants are fixed in part by the desired
boundary conditions, and in part by the equations of motion outside the core. 
In the exterior region, the O($E^{-2}$) terms require
\be
q_{0M} - d_0 + f_0 = D(\infty) \;\;\; {\rm or} \;\;\;
q_{0E} - d_0 - f_0 = 0
\ee 
However, note that if $a\neq-1$, then $f(\sigma)$ grows logarithmically,
and eventually, the simple form of the dilaton no longer satisfies
the equations of motion, and instead will have a more complicated form.
This is a feature of the strong asymptotic effect of the vortex on the dilaton
in the very far field regions already seen in the self-gravitating
dilatonic vortex \cite{GS}. 
It  will however only happen at a large length scale, and only for a massless
dilaton. This suggests that $a\neq -1$ vortices are unsuitable for using to 
smooth out conical deficits.

To make this more precise, consider how one might smooth out the conical
deficit of the dilatonic C-metric. Since the metric already has an effective
asymptotic deficit angle, and a conical singularity where we wish to place
the core of the string, the appropriate boundary conditions are that the
perturbations rapidly vanish outside the core of the vortex, that is, for
$a = -1$ we choose 
\be
b_0 = A(\infty) \;,\;\;\; d_0 = f_0 = -{\half} D(\infty) = {\half}g_0 \;,
\;\;\; q_{0M} = q_{0E}=0
\ee
If we examine the condition for elementary flatness of the metric 
at the core of the vortex, we obtain $\delta\varphi = 2\pi\epsilon(b_0-g_0)
= \epsilon \mu$ as required. However, note that if $a\neq-1$, the dilaton
has a logarithmic divergence at large scales, and we can never have the 
perturbations vanishing outside the core. We therefore conclude that in 
this case the vortex cannot be used to smooth out the singularities
in the dilatonic C-metrics. 

Focussing on the string threading the black hole, and transforming back
to spherically symmetric coordinates (\ref{sphsym}), we see that
in general, the geometry is corrected to
\be
ds^2 = \left ( 1 + D e^{2(a+1)\phi_0} \right ) \left [ 
\lambda dt^2 - \lambda^{-1} dr^2
- C^2d\theta^2 - C^2 \left (1-{\epsilon\mu\over2\pi} 
e^{2(a+1)\phi_0} \right)\sin^{2}\theta d\varphi^2 \right ]
\ee
at least for some intermediate range of $r$, and hence is not a simple conical
deficit. For example, the black hole with a massless dilaton outside the
vortex core becomes
\bsq\bq
ds^2 &=& \left ( 1 + D \left ( 
\textstyle{ 1 - {Q^2\over Er}} \right )^{-(a+1)} \right ) \times 
\Biggl [ \left ( \textstyle{ 1 - {2E\over r}} \right ) dt^2  \nonumber \\
&& - \left ( \textstyle{ 1 - {2E\over r}} \right ) ^{-1} dr^2 
- r\left ( \textstyle { r-{Q^2\over E}} \right ) \left ( d\theta^2 
+ \left ( \textstyle{ 1 - {\epsilon\mu\over2\pi}\left ( 1 - {Q^2\over Er} 
\right )^{-(a+1)}}\right )^2 d\varphi^2 \right ) \Biggr ] \\
e^{\pm 2\phi} &=& \left ( 1 - {Q^2\over Er} \right ) \left (
1 \pm 2 \epsilon f(\sigma) \left ( 1 - {Q^2\over Er} \right )^{\pm(a+1)}
\right ) \label{cordil}\\
A_{\nu} &=& \cases{ {Q\over r} \; \partial_\nu t & electric, \cr
Q (1-\cos\theta) [1-\epsilon(A_\infty +D_\infty 
- 2(a+1)\mu \ln(r\sin\theta) )]\; \partial_\nu \phi & magnetic.\cr}
\eq\esq
where the two roots in (\ref{cordil})\ correspond to electric
and magnetic black holes respectively.
This allows us to quantify precisely the limits of validity of our
approximation. If $a\neq-1$, then it is easy to see that at very large
distances, the strong effect of the vortex on the dilaton means that
our simple form of the perturbation is no longer valid.  
Across the horizon there is  an additional dilaton flux 
switched on, and we see that in spite of the fact that the thin vortex
solution works for an extremal magnetic black hole, the back reaction
for $a>-1$ is badly behaved at the horizon. 

For $a = -1$, none of these problems arise, and we simply have a gentle
shift in the value of the dilaton generated by the radial stresses of
the vortex, $\phi(\infty) \to \phi(\infty) - \half\epsilon D(\infty)$,
which can be either positive, negative or zero depending on whether
$\beta$ is greater than, less than, or equal to unity. Note that this
shift has the same sign for both magnetic and electric black holes, so
that if the dilaton is increased in magnitude for an electric black hole,
it is decreased in magnitude for the magnetic one, and vice versa.
For $\beta = 1$, the only fields affected by the vortex are the 
$g_{\varphi\varphi}$ component of the metric, and the magnetic
potential. This is the true model vortex for smoothing a conical 
singularity.

\section{\bf Conclusions.}\label{finsec}
 
To summarize, we have provided analytic arguments to show that a vortex
can sit on a black hole horizon in dilatonic gravity, much the same as
in Einstein gravity, the crucial difference being that for near extremal
electrically charged black holes, the thin vortex approximation ceases
to hold, and the flux starts to expel, however, this can be viewed as a
consequence of the vanishing area of the horizon. 
For the case of massive dilatonic black holes, the thin vortex approximation
was shown to hold in a range of cases, the only exception being near
extremal electrically charged black holes for a small dilaton mass.
We should also point out that these arguments can be used to paint a global 
vortex onto the dilatonic black hole, since a global vortex is obtained
by setting $P=1$, $\beta \to \infty$. However, in this case, we might expect
the gravitational back reaction to be problematic, given the nature
of the Einstein global string metric, which is not only non-asymptotically
locally flat, but also time dependent \cite{global}.

For extremal black holes, we were able to prove analytically that flux
expulsion occurs for all extremal magnetically charged black holes, 
independent of the mass of the dilaton, and for electrically charged
extremal black holes if $QM>e/2$. If $QM<e/2$, and the black hole is
extremal and electric, then it will have vanishing area, and in some
sense the flux is trivially expelled. This phenomenon of flux expulsion
is quite generic, and occurs because the equations of motion on the horizon
decouple from the exterior, thus forcing the vortex to sit to a compact  
subspace of the full spacetime. This phase transition in the existence 
of topologically nontrivial field configurations on compact spaces as
a function of the size of that space has been observed for example in
domain walls \cite{walls}, where the self-gravitation of the wall 
is responsible for the compactification of the spatial sections. It is
also observed in pure electromagnetism (i.e.\ in the absence of any
broken symmetry and topological defects) \cite{emexp}. What our work has
shown is that the presence of the dilaton does not destroy this phenomenon,
as indeed one would expect if it were directly related to the 
area of a compact subspace - the horizon.

The gravitational back reaction of the vortex was found assuming the
validity of thin vortex approximation. The spacetime was found to be
approximately conical to leading order, however, if the dilaton is
massless, and if $a\neq-1$, there are long range effects on the geometry,
which is not precisely conical. For $a = -1$, the fields are well behaved,
and the vortex can be used to smooth out the conical singularities of the
dilatonic C-metrics for example. Except for the very special case of
$\beta = 1$, $a = -1$, in which case the vortex couples only to the geometry
and not to the dilaton, the presence of the vortex breaks the electromagnetic
duality invariance of the equations of motion, although only at the 
O($\epsilon$) level.

Finally, one criticism of this method might be that although we have not
been completely restrictive in our choice of coupling of the abelian-Higgs
model to string gravity, in that we included an arbitrary coupling in
the string frame $a$, we have not been entirely general either, in that
we could have had the abelian-Higgs lagrangian, (\ref{abhiggs}), coupled
by an arbitrary parameter in an arbitrary frame. This obviously 
increases the complexity of the analysis, without adding any great 
additional insight, therefore we chose not to include this additional
complication while deriving the main results, however, we would now like
to conclude by indicating how our results are modified if this most
general case scenario is considered.

Let us suppose the dilaton couples in a frame ${\tilde{\bf F}}$,
which is related to the string frame via
\be
{\hat g}_{ab} = e^{2b\phi} {\tilde g}_{ab}
\ee
so that for example, $b=1$ for the Einstein frame.
Then the string energy-momentum tensor (\ref{stringem}) in the Einstein
frame is modified to
\bq
{\hat T}_{ab} &=& 2
e^{2(a-b+1)\phi} \left[ \nabla_aX\nabla_bX + X^2 P_aP_b \right] - \beta
e^{2a\phi} G_{ac}G_b^{\ c} \nonumber \\ && -
e^{2(a-b+1)\phi} g_{ab} \left [ (\nabla X)^2 + X^2 P_a^2 - {\beta\over2}
e^{2(b-1)\phi} G^2 - {\quarter} e^{2(1-b)\phi} (X^2-1)^2 \right ] 
\eq
While this does not affect the spacetime geometry of the self-gravitating
dilatonic vortex, it does modify the dilaton solution, changing the 
$(a+1)$ factor in front of ${\cal E}_0$ to $(a+1-b)$, and the `$\half$' in 
front of $({\cal P}_{_0R} + {\cal P}_{_0\varphi})$ to ${1-b\over2}$, so
that, for example, the massless dilaton asymptotes
\be\label{newmsls}
\phi \sim (a+1-b) {\epsilon \mu\over 2\pi} \ln R
- {\epsilon(1-b)\over2} D(\infty) \;\;\;{\rm as} \;\; R\to\infty
\ee

For the arguments of section~\ref{thinsec}\ we now find that 
$\sigma = Ce^{(1-b)\phi}\sin\theta$, and the correcting terms 
${\cal T}_i$ defined by (\ref{corrterms}) should be modified by
replacing $Ce^\phi$ by $Ce^{(1-b)\phi}$ whenever it appears. 
It is a somewhat long but straightforward calculation to show that
the conclusions of this section are not modified except near
the extremal limit. For example, for the magnetic black hole 
\bsq\bq
{\cal T}_1 &=& {\sigma^2\over r^2} \left ( 1 - {Q^2\over Er} \right )^{-b}
{2E\over r}\left [
1 + (1-b) \left ({Q^2 \over 2E^2} - {Q^2 \over Er} \right )
- {b^2Q^4 \over 8E^2r} {(r-2E)\over(r-Q^2/E)} \right ] \\
{\cal T}_2 &=& {\sigma^2\over r^2} \left ( 1 - {Q^2\over Er} \right )^{-b}
\Biggl [ {2E\over r}
\left (1 - {Q^2 \over Er} \right ) - {aQ^2 \over Er}
\left ( 1 - {2E\over r} \right ) \nonumber \\
&& + {bQ^2\over r^2} \left [
1 + (b-2-2a) {Q^4(r-2E) \over 4E^2(r-Q^2/E)} \right ] \Biggr ] 
\eq\esq
Clearly, unless $b>0$, these terms are always O($E^{-2}$), however,
if $b>0$, then the extremal magnetic black hole no longer admits a
thin vortex limit. It is easy to see why this is, since in the frame
${\tilde{\bf F}}$, the extremal metric is now
\be
{\tilde ds}^2 = \left ( 1 - {2E\over r} \right )^b \left [
dt^2 - {dr^2 \over \left ( 1 - {2E\over r} \right )^2}
- r^2 d\theta^2 \ r^2 \sin^2\theta d\varphi^2 \right ]
\ee
in which the horizon now has vanishing area for positive $b$, hence
the thin vortex approximation is breaking down for the same reason
as in the case of the extremal electric black hole. Similarly, for
extremally charged electric black holes, we now obtain that
the thin vortex approximation always works if $b\leq -2$.

Finally, for the gravitational back reaction (when the thin vortex
approximation can be used) the form of the corrections to the geometry and
dilaton will now be modified to
$ \alpha = \alpha_0 \left ( 1 + \epsilon e^{2(a+1-b)\phi_0} b(\sigma)\right )$
etc., where the same functions $f$, $d$, $g$ are used, $\sigma$ is
the one appropriate to the new frame as defined above, and $f$ is now the
new dilaton function, defined above for the massless dilaton in (\ref{newmsls}).
The main alterations of the conclusions of this section are that it is now
for $a = b-1$ (rather than for $a = -1$) that 
we have a good far field behaviour for the perturbed
geometry and dilaton functions, and we no longer require a Bogomolnyi
vortex for the dilaton to be unaffected; we can also choose to set $b=1$,
(and $a=0$) i.e., to couple the string in the Einstein frame.

\section*{\bf Acknowledgements.}

We would like to thank  Filipe Bonjour and Roberto Emparan
for useful discussions, and Rafal Moderski and Marek Rogatko for
making available their results on analysing extremal electric black
holes.  This work was supported by an FCT - Portugal fellowship BD/5814/95
(CS) and by the Royal Society (RG).


\begin{thebibliography}{88}
\def\apj#1 #2 #3.{{\it Astrophys.\ J.\ \bf#1} #2 (#3).}
\def\cmp#1 #2 #3.{{\it Commun.\ Math.\ Phys.\ \bf#1} #2 (#3).}
\def\comnpp#1 #2 #3.{{\it Comm.\ Nucl.\ Part.\ Phys.\  \bf#1} #2 (#3).}
\def\cqg#1 #2 #3.{{\it Class.\ Quant.\ Grav.\ \bf#1} #2 (#3).}
\def\grg#1 #2 #3.{{\it Gen.\ Rel.\ Grav.\ \bf#1} #2 (#3).}
\def\jmp#1 #2 #3.{{\it J.\ Math.\ Phys.\ \bf#1} #2 (#3).}
\def\mpla#1 #2 #3.{{\it Mod.\ Phys.\ Lett.\ \rm A\bf#1} #2 (#3).}
\def\ncim#1 #2 #3.{{\it Nuovo Cim.\ \bf#1\/} #2 (#3).}
\def\npb#1 #2 #3.{{\it Nucl.\ Phys.\ \rm B\bf#1} #2 (#3).}
\def\phrep#1 #2 #3.{{\it Phys.\ Rep.\ \bf#1\/} #2 (#3).}
\def\plb#1 #2 #3.{{\it Phys.\ Lett.\ \bf#1\/}B #2 (#3).}
\def\pr#1 #2 #3.{{\it Phys.\ Rev.\ \bf#1} #2 (#3).}
\def\prd#1 #2 #3.{{\it Phys.\ Rev.\ \rm D\bf#1} #2 (#3).}
\def\prl#1 #2 #3.{{\it Phys.\ Rev.\ Lett.\ \bf#1} #2 (#3).}
\def\pt#1 #2 #3.{{\it Phys.\ Today.\ \bf B#1} #2 (#3).}

 
\bibitem{RW} R.Ruffini and J.A.Wheeler, \pt 24 30 1971.

\bibitem{5Chrusciel} 
R.Bartnik and J.McKinnon, \prl 61 141 1988.\hfill\break
K.Lee, V.Nair and E.Weinberg, \prl 68 1100 1992. [hep-th/9111045]. 
\prd 45 2751 1992. [hep-th/9112008].\hfill\break
F.Dowker, R.Gregory and J.Traschen, \prd 45 2762 1992. [hep-th/9112065].

\bibitem{AGK} A.Ach\'ucarro, R.Gregory
and K.Kuijken, \prd 52 5729 1995. [gr-qc/9505039].

\bibitem{nohair}
S.L.Adler and R.B.Pearson, \prd 18 2798 1978.\hfill\break
A.Lahiri, \mpla 8 1549 1993. [gr-qc/9207008].\hfill\break
P.T.Chru\'sciel, {\it No hair theorems - Folklore, Conjectures, Results},
gr-qc/9402032. \hfill\break
J.D.Bekenstein, \prd 51 6608 1995.\hfill\break
A.E.Mayo and J.D.Bekenstein, \prd 54 5059 1996. [gr-qc/9602057].

\bibitem{split} 
S.W.Hawking and S.F.Ross,  \prl 75 3382 1995. [gr-qc/9506020].\hfill\break
R.Emparan,  \prl 75 3386 1995. [gr-qc/9506025].\hfill\break
D.Eardley, G.Horowitz, D.Kastor and J.Traschen,
\prl 75 3390 1995. [gr-qc/9506041].

\bibitem{GH} R.Gregory and M.Hindmarsh, Phys. Rev. D. 52, 5598 1995.
[gr-qc/9506054].

\bibitem{E2} R.Emparan, \prd 52 6976 1995. [gr-qc/9507002].

\bibitem{AFV} M.Aryal, L.Ford, and A.Vilenkin, \prd 34 2263 1986.

\bibitem{KW} W.Kinnersley and M.Walker, \prd 2 1359 1970.

\bibitem{IK} W.Israel and K.A.Khan, \ncim 33 331 1964.

\bibitem{CCES} A.Chamblin, J.Ashbourn-Chamblin,
R.Emparan and A.Sornborger, \prl 80 4378 1998. [gr-qc/9706032].
\prd 58 124014 1998. [gr-qc/9706004].

\bibitem{BG} F.Bonjour and R.Gregory, \prl 81 5034 1998. [hep-th/9809029].

\bibitem{BRG} F.Bonjour, R.Emparan and R.Gregory, \prd 59 084022 1999. 
[gr-qc/9810061].

\bibitem{MR} R.Moderski and M.Rogatko, \prd 58 124016 1998. [hep-th/9808110].

\bibitem{RJ} R.Gregory and J.Harvey, \prd 47 2411 1993. [hep-th/9209070]

\bibitem{HH} J.H.Horne and G.Horowitz, \npb 399 169 1993. [hep-th/9210012]

\bibitem{GHS} G.W.Gibbons and K.I.Maeda, \npb 298 741 1988. \hfill\break
D.Garfinkle, G.Horowitz and A.Strominger, \prd 43 3140 1991.

\bibitem{GG} D.Garfinkle, \prd 32 1323 1985. R.Gregory, \prl 59 740 1987.

\bibitem{GS} R.Gregory and C.Santos, \prd 56 1194 1997. [gr-qc/9701014].

\bibitem{LESG} E.Fradkin and A.Tseytlin, \plb 158 316 1985. \hfill\break
C.Callan, D.Friedan, E.Martinec and M.Perry, \npb 262 593 1985. \hfill\break
C.Lovelace, \npb 273 413 1986.

\bibitem{bog} E.B.Bogomolnyi, {\it Yad.~Fiz.} {\bf24} 861 (1976).
[{\it Sov.~J.~Nucl.~Phys.~{\bf 24}} 449 (1976)]

\bibitem{NO} H.Nielsen and P.Olesen, \npb 61 45 1973.

\bibitem{AGHK}  A.Ach\'ucarro, R.Gregory, J.Harvey and K.Kuijken, 
\prl 72 3646 1994. [hep-th/9312034].

\bibitem{VHL} A.Vilenkin, \prd 23 852 1981.
J.R.Gott III, \apj 288 422 1985.\hfill\break
W.Hiscock, \prd 31 3288 1985.
B.Linet, \grg 17 1109 1985.

\bibitem{MRpc} R.Moderski and M.Rogatko, private communication.

\bibitem{DGKT} F.Dowker, J.Gauntlett, D.Kastor and J.Traschen, \prd 49 2909
1994. [hep-th/9309075].

\bibitem{global} R.Gregory, \prd 54 4955 1996. [gr-qc/9606002]

\bibitem{walls} R.Basu and A.Vilenkin, \prd 46 2345 1992. \prd 50 7150 1994.
[gr-qc/9402040] \hfill\break
F.Bonjour, C.Charmousis and R.Gregory, \cqg 16 2427 1999. [gr-qc/9902081],
and gr-qc/9903059.

\bibitem{emexp} A.Chamblin, R.Emparan and G.Gibbons, \prd 58 084009 1998.
[hep-th/9806017].

\end{thebibliography}
\end {document}